\documentclass[12pt]{article}

\catcode`\@=11
\@addtoreset{equation}{section}

\global\arraycolsep=2pt
\usepackage{amsbsy,amssymb,latexsym,amsfonts,amsmath}
\usepackage{graphicx,color}

\allowdisplaybreaks

\begin{document}
\begin{flushright}
\parbox{4.2cm}
{UCB-PTH-09/29}
\end{flushright}

\vspace*{0.7cm}

\begin{center}
{\Large \bf 
No Forbidden Landscape in String/M-theory}
\vspace*{2.0cm}\\
{Yu Nakayama}
\end{center}
\vspace*{-0.2cm}
\begin{center}
{\it Berkeley Center for Theoretical Physics, \\ 
University of California, Berkeley, CA 94720, USA
}
\vspace{3.8cm}
\end{center}

\begin{abstract} 
Scale invariant but non-conformal field theories are forbidden in $(1+1)$ dimension, and so should be the corresponding holographic dual gravity theories.  
We conjecture that such scale invariant but non-conformal field configurations do not exist in the string/M-theory. We provide a proof of this conjecture in the classical supergravity limit under a certain gauge condition. Our proof does also apply in higher dimensional scale invariant but non-conformal field configurations, which suggests that scale invariant but non-conformal field theories may be forbidden in higher dimensions as well.

\end{abstract}

\thispagestyle{empty} 

\setcounter{page}{0}

\newpage

\section{Introduction} 
It is one of the outstanding problems in theoretical physics to map a distribution of landscape and swampland \cite{Vafa:2005ui}\cite{Ooguri:2006in} and draw accurate lines between them.
 Careful studies of the consistency in quantum theories of gravity have revealed that not all the effective quantum field theories can possess an ultra-violet completed description consistent with the quantum gravity. Still, it is extremely difficult to locate whether a given theory is inside or outside of the swampland.

It has been suggested that the holography \cite{hol1}\cite{hol2} is one of the fundamental principles of quantum theories of gravity, and the (in)consistency of the dual field theory would yield strong constraints on the properties of the quantum theories of gravity. Recently in a beautiful paper by Hellerman \cite{Hellerman:2009bu}, a strong constraint on the mass spectrum of any consistent quantum gravity with $AdS_3$ background has been uncovered. The constraint comes from the unitarity and the modular invariance of the dual boundary field theory.

With the same spirit but from a completely different viewpoint, in \cite{Nakayama:2009qu} it has been shown that a certain class of effective field theories coupled with the Einstein gravity would produce inconsistent field configurations that can never occur from the dual field theory perspective. As a consequence, \cite{Nakayama:2009qu} has concluded that such effective field theories must be forbidden in any consistent theories of quantum gravity.
Here, the constraint comes from the theorem proved by Polchinski \cite{Polchinski:1987dy}: scale invariant field theories must be conformally invariant under a few mild assumptions in $(1+1)$ dimension.\footnote{The ``forbidden landscape" refers to such seemingly good effective field theories but inconsistent with the holographic interpretation once coupled to gravity. In this paper, we would like to give a partial evidence how such theories cannot occur in the string/M-theory, so we have ``no" forbidden landscape from the viewpoint of Polchinski's theorem.} 

Gravitational counterpart of Polchinski's theorem claims that consistent quantum theories of gravity may never produce such inconsistent field configurations dual to scale invariant but non-conformal field theories, and the requirement that such solutions should not exist will constrain possible effective field theories that can be embedded in ultra-violet completed quantum theories of gravity. The argument is strong in the sense that we do not assume any microscopic quantum theories of gravity such as the string/M-theory. Rather we have to test whether the string/M-theory satisfies the constraint from Polchinski's theorem once we assume the validity of the holography.  
 
This paper is devoted to this task. We formulate the gravitational counterpart of Polchinski's theorem for string/M-theory (which can be generalized to other effective field theories that one believes to appear in the low energy limit of the consistent quantum theories of gravity she/he likes). Since we believe that the string/M-theory is a consistent theory of quantum gravity, we naturally conjecture that the field configurations dual to  scale invariant but non-conformal field theories in $(1+1)$ dimension never appear in the string/M-theory compactification. We give evidence of this conjecture by studying the low-energy supergravity background with the most general scale invariant ansatz.

The conjecture further gives us a restriction on the structure of higher derivative/loop corrections to supergravity limit of the string/M-theory action. In \cite{Nakayama:2009qu}, a particular class of effective actions (so called ``spontaneous Lorentz symmetry breaking model" or ``(gauged) ghost condensation model") are excluded precisely because they predict the scale invariant but non-conformal field configurations once they are coupled to the Einstein gravity. Similarly, higher-derivative/loop corrections in the string/M-theory should never introduce such a structure so as not for the scale invariant but non-conformal field configurations  to be possible.

\section{Conjecture}
Let us first state  our conjecture in the M-theory compactification. As we will see shortly, similar statements also apply in the string compactification. The low energy limit of the M-theory is described by the (1+10) dimensional supergravity with the bosonic action
\begin{align}
 S_{M}= -\frac{1}{2\kappa_{11}^2} \int d^{11} x \sqrt{-g} \left(R- \frac{1}{2\cdot 4 !} F_{\mu\nu\rho\sigma} F^{\mu\nu\rho\sigma} \right) - \frac{1}{6}\int C_3 \wedge F_{4} \wedge F_4 \ ,
\end{align}
where $F_4 = dC_3 = F_{\mu\nu\rho\sigma} dx^{\mu}dx^\nu dx^{\rho} dx^{\sigma}$. The conjecture excludes a certain class of field configurations in the M-theory background.

{\bf Conjecture (in M-theory)}

The most general M-theory background ansatz dual to a $(1+1)$ dimensional field theory invariant under the scale transformation\footnote{We always assume $(1+1)$ dimensional Poincar\'e invariance in the following discussions.} is given by
\begin{align}
ds^2_{11} = f(\xi) \frac{dz^2}{z^2} + g(\xi)\frac{-dt^2+dx^2}{z^2} + 2h_i(\xi) \frac{dz d\xi^i}{z} + ds^{2}_8(\xi) \ 
\end{align}
where $ds^{2}_8(\xi)$ is the metric for the internal eight-fold $X_8$ and $i$ runs from $2$ to $10$. The scale transformation acts as
\begin{align}
z \to \lambda z \ , \ \ t \to \lambda t \ , \ \ x \to \lambda x \ , \ \ 
\end{align}
The most general ansatz for the four-form flux is given by
\begin{align}
F_4 = A + \frac{dz}{z} \wedge B + \frac{dt\wedge dx}{z^2} \wedge C + \frac{dz\wedge dt \wedge dx}{z^3} \wedge D  \ ,
\end{align}
where $A$, $B$, $C$ and $D$ are differential forms on the internal manifold $X_8$ and do not depend on $(z,t,x)$.
 
The conjecture is that all the M-theory solutions dual to scale invariant $(1+1)$ dimensional field theories have a larger $AdS_3$ isometry and invariant flux. In other words, $f(\xi) = g(\xi)$, and $h_{i}(\xi) = B=C= 0$ up to a diffeomorphism. If and only if these conditions are satisfied, the geometry is invariant under the special conformal transformation:
\begin{align}
\delta x_a = 2(\epsilon^{b}x_b)x_a - (z^2 + x^bx_b)\epsilon_a \ , \ \ \delta z = 2(\epsilon^{b}x_b)z
\end{align}
A typical solution of this form is $AdS_3 \times S_1 \times T^4 \times S^3$, but we claim that the condition should be satisfied by all the consistent solutions of the M-theory equation of motion, even with higher derivative/loop corrections.

{\bf Conjecture (in string theory)}

A similar conjecture naturally arises in the string compactification as well.
The type IIA string theory has the following bosonic effective action
\begin{align}
S_{IIA} &= \frac{1}{2\kappa^2} \int d^{10} x \sqrt{-g} e^{-2\Phi} \left(R + 4\partial_\mu \Phi \partial^\mu \Phi - \frac{1}{2}|H_{3}|^2 \right) \cr
  & - \frac{1}{4\kappa^2} \int d^{10} x \sqrt{-g}\left(|F_{2}|^2 + |\tilde{F}_{4}|^2 \right) -\frac{1}{4\kappa^2} \int B_{2} \wedge F_{4} \wedge F_{4} \ , 
\end{align}
where $ \tilde{F}_{4} = dA_{3} + A_{1} \wedge H_{3}$, and $H_3 = dB_2$.

In the type IIA string theory, the most general background ansatz dual to a $(1+1)$ dimensional field theory invariant under the scale transformation is given by
\begin{align}
ds^2_{10} = f(\xi) \frac{dz^2}{z^2} + g(\xi)\frac{-dt^2+dx^2}{z^2} + 2h_i(\xi) \frac{dz d\xi^i}{z} + ds^{2}_7(\xi) \ ,
\end{align}
where $ds^{2}_7(\xi)$ denotes the metric for the internal seven-fold $X_7$. 
The most general ansatz for the flux is given by
\begin{align}
F_{2} = A_{(2)} + \frac{dz}{z} \wedge B_{(2)} + \frac{dt \wedge dx} {z^2} C_{(2)} \cr
H_{3} = A_{(3)} + \frac{dz}{z} \wedge B_{(3)} + \frac{dt\wedge dx}{z^2} \wedge C_{(3)} + \frac{dz \wedge dt \wedge dx}{z^3} \wedge D_{(3)}  \cr
\tilde{F}_{4} = A_{(4)} + \frac{dz}{z} \wedge B_{(4)} + \frac{dt\wedge dx}{z^2} \wedge C_{(4)} + \frac{dz\wedge dt \wedge dx}{z^3} \wedge D_{(4)}  \ , 
\end{align}
where $A_{(a)}$, $B_{(a)}$, $C_{(a)}$ and $D_{(a)}$ are differential forms on the internal manifold $X_7$ and do not depend on $(z,t,x)$. The dilaton may depend on $\xi^i$ but does not depend on $(z,t,x)$.

The conjecture is that all the consistent solutions dual to scale invariant $(1+1)$ dimensional field theories have a larger $AdS_3$ isometry and invariant flux. In other words, $f(\xi) = g(\xi)$, and $h_{i}(\xi) = B_{(a)} = C_{(a)} = 0 $ up to a diffeomorphism. 
A typical example of such a solution is $AdS_3 \times T^4 \times S_3$, but again we claim that the condition should be satisfied by all the consistent solutions of the type IIA string theory, even with higher derivative/loop corrections.

The type IIB string theory has the following effective action
\begin{align}
S_{IIB} &= \frac{1}{2\kappa^2} \int d^{10} x \sqrt{-g} e^{-2\Phi} \left(R + 4\partial_\mu \Phi \partial^\mu \Phi - \frac{1}{2}|H_{3}|^2 \right) \cr
  & - \frac{1}{4\kappa^2} \int d^{10} x \sqrt{-g}\left(|F_{1}|^2 + |\tilde{F}_{3}|^2 +\frac{1}{2}|\tilde{F}_5|^2 \right) -\frac{1}{4\kappa^2} \int C_{4} \wedge H_{3} \wedge F_{3} \ , 
\end{align}
where $\tilde{F}_3 = F_3-C_0H_3$ and $\tilde{F}_5 = F_5 -\frac{1}{2}C_2\wedge H_3 +\frac{1}{2}B_2\wedge F_3$. We will impose the self-duality condition $\tilde{F}_5 = \star \tilde{F}_5$ by hand after deriving the equations of motion.

In the type IIB string theory, the most general background ansatz dual to a $(1+1)$ dimensional field theory invariant under the scale transformation is given by
\begin{align}
ds^2_{10} = f(\xi) \frac{dz^2}{z^2} + g(\xi)\frac{-dt^2+dx^2}{z^2} + 2h_i(\xi) \frac{dz d\xi^i}{z} + ds^{2}_7(\xi) \ .
\end{align}
The most general ansatz for the flux ($G_3 = F_3 - \tau H_3$, $\tau = C_0 + ie^{-\Phi}$) is given by
\begin{align}
G_3 = A_{(3)} + \frac{dz}{z} \wedge B_{(3)} + \frac{dt\wedge dx}{z^2} \wedge C_{(3)} + \frac{dz\wedge dt \wedge dx}{z^3} \wedge D_{(3)}  \cr
\tilde{F}_5 = A_{(5)} + \frac{dz}{z} \wedge B_{(5)} + \frac{dt\wedge dx}{z^2} \wedge C_{(5)} + \frac{dz\wedge dt \wedge dx}{z^3} \wedge D_{(5)}  \ ,
\end{align}
where $A_{(a)}$, $B_{(a)}$, $C_{(a)}$, and $D_{(a)}$ are differential forms on the internal manifold $X_7$ and do not depend on $(z,t,x)$. The axio-dilaton $\tau$ does not depend on $(z,t,x)$ but may depend on $\xi^i$.

The conjecture is that all the consistent solutions dual to scale invariant $(1+1)$ dimensional field theories have a larger $AdS_3$ isometry and invariant flux. In other words, $f(\xi) = g(\xi)$, $h_{i}(\xi) = B_{(a)}= C_{(a)} = 0 $ up to diffeomorphism invariance. 
The conjectures can be generalized to the type I or heterotic string theory because the supersymmetry does not play any role in our discussion.

{\bf Holographic origin of the conjecture}

The conjectures above, both in the M-theory and in the string theories, are based on the holography. As stated in the introduction, Polchinski proved the theorem that the scale invariant field theories in $(1+1)$ dimension are necessarily conformally invariant under the assumptions \cite{Polchinski:1987dy}:
\begin{itemize}
	\item the theory is unitary
	\item the theory is Poincar\'e invariant, and
	\item the theory has a discrete spectrum.
\end{itemize}
All these assumptions are reasonable to make in the compactification we are investigating.\footnote{Once these assumptions are relaxed, there are known examples of scale invariant but non-conformal field theories (see e.g. \cite{Hull:1985rc}\cite{Riva:2005gd}\cite{Ho:2008nr}).} Thus as long as the holography is correct, and the string/M-theory is consistent, the scale invariant field configurations of the string/M-theory should be necessarily conformally invariant. 

It would be interesting to see what happens when some of the above assumptions are not met in the gravity side. For example, in \cite{Awad:2000ie}, the example of non-conformal but scale invariant geometry was presented by placing the dual field theory on non-trivial background. The non-unitary examples might be easily found by flipping some of the sign of the flux kinetic terms as we will see. In this paper, however, we would strictly restrict ourselves to theories with original assumptions made in \cite{Polchinski:1987dy}.

This gravitational counterpart of Polchinski's theorem has led to the conjectures in this section.  We will see the evidence of the conjectures in the supergravity limit in the next section. 
The higher derivative/loop corrections should conspire themselves so that such solutions cannot be realized as a solution of the equation of motion.\footnote{We set the classical value of the fermionic fields all zero in the discussion of the supergravity limit below, but the conjecture itself does not exclude the possible quantum condensation of the fermionic fields as long as they preserve the full conformal invariance. The condensation that preserves the scale invariance but not the special conformal invariance should be forbidden.} We will discuss the constraint on the higher derivative/loop corrections from this point of view in section 5.

\section{Evidence in the supergravity limit}
In this section, we verify the conjectures presented in the last section in the supergravity limit. The argument in this section applies both to the string theories and the M-theory, so we will treat them in a uniform manner. Let us begin with the $ D (=10,11)$ dimensional Einstein gravity coupled with several form fields:
\begin{eqnarray}
 S = \int d^D x \sqrt{-g} \left(\frac{1}{2\kappa^2} R + \sum_a |F_{(a)}|^2\right)   +  (\text{CS terms}) \ , \label{caction}
\end{eqnarray}
where the suffix $a$ distinguish all the form fields of the theory under consideration, and $|F_{(a)}|^2$ is a schematic way to denote their kinetic terms.
The assumption (satisfied in the string/M-theory) here is that the non-trivial interaction terms are encoded in the Chern-Simons terms so that they do not contribute to the energy momentum tensor. Since we are using the Einstein frame, the Bianchi identities and equations of motion for flux may become complicated, but we need not use them in the following arguments.

The Einstein equation from the action \eqref{caction} takes the following form
\begin{eqnarray}
 R_{\mu\nu} = \sum_{a} -c_a|F_{(a)}|^2 g_{\mu\nu} + c'_a F_{(a) \mu \rho \cdots}F_{(a) \nu}^{\ \ \ \rho \cdots} \ ,
\end{eqnarray}
where $c_a$ and $c'_{a}$ are positive constants that depend on the degree of the forms and the dimensionality of the space-time. As long as they are positive, they are not important in the following argument.

The most general background ansatz dual to a $(1+1)$ dimensional field theory invariant under the scale transformation is given by
\begin{align}
ds^2_{D} = f(\xi) \frac{dz^2}{z^2} + g(\xi)\frac{-dt^2+dx^2}{z^2} +2h_i(\xi) \frac{dz d\xi^i}{z} + ds^{2}_{D-3}(\xi) \ . \label{aaab}
\end{align}
The most general ansatz for the flux is given by
\begin{align}
F_{(a)} = A_{(a)} + \frac{dz}{z} \wedge B_{(a)} + \frac{dt\wedge dx}{z^2} \wedge C_{(a)} + \frac{dz\wedge dt \wedge dx}{z^3} \wedge D_{(a)}
\end{align}

As a warm up, let us begin with the higher dimensional analogue of the background studied in \cite{Nakayama:2009qu}, where $h_{i}(\xi) = 0$ and $f(\xi) = g(\xi)$. The metric is the warped product of $AdS_3$ and the internal $X_{D-3}$, and the special conformal invariance would be broken only in the matter (flux) sector. This would be a good approximation of the flux compactification when the internal space is small compared with the curvature radius of the non-compact space (and hence the Kaluza-Klein reduction is trustful).

The symmetry of the metric tells us that $R_{zz} + R_{tt} = 0$, while the Einstein equation tells us that it is equal to the sum of the flux bilinear:
\begin{align}
 0 = R_{zz} + R_{tt}  = \frac{1}{z^2}\sum_a c'_a|B_{(a)}|^2 + c'_a|C_{(a)}|^2 \ .
\end{align}
Since the right hand side is the sum of semi-positive definite flux bilinears, they must vanish separately: $B_{(a)} = C_{(a)}  = 0$, which completes the proof: scale invariance is enhanced to the conformal invariance due to the Einstein equation.

Let us move on to the most general situation where we can give a proof. For this purpose, we impose the gauge condition $h_i(\xi) = f(\xi) \partial_i\Lambda(\xi)$. Then, it is always possible to gauge away $h_i(\xi)$ by the diffeomorphism invariance. The metric takes the form of \eqref{aaab} without $h_i$. We compute the combination of the Ricci tensor $g(\xi)R_{zz} + f(\xi) R_{tt}$
\begin{align}
 \frac{\alpha z^2}{f^{\frac{1}{2}}} \left( g(\xi) R_{zz} + f(\xi) R_{tt} \right) = D^i \left( -2f^{-\frac{1}{2}} g \partial_i f + 2 f^{\frac{1}{2}} \partial_i g \right) \ , \label{totall}
\end{align}
which is proportional to a total derivative. Here $D_i$ is the covariant derivative constructed from the internal metric $ds^{2}_{D-3}$, and $\alpha$ is a positive constant, which depends on the dimensionality of the space-time. Since it is a total derivative, integral of \eqref{totall} over the internal space $X_{D-3}$ must vanish. 

It is well-known that the 10 or 11 dimensional supergravity action satisfies the null energy condition \cite{No}, so the left hand side of \eqref{totall} is semi-positive definite. More precisely the Einstein equations tell us that the left hand side is the sum of semi-positive definite flux bilinears: $\sum_a c'_a|B_{(a)}|^2 + c'_a|C_{(a)}|^2 = 0$, which demands $B_{(a)} = C_{(a)} = 0$ after integrating \eqref{totall} over $X_{D-3}$. The $R_{zi}$ component of Einstein's equation then gives $f(\xi) = g(\xi)$ up to a scaling factor that can be absorbed by the redefinition of $x$ and $t$.
 Therefore, we have verified that in the supergravity limit under a certain gauge condition, the scale invariant field configurations automatically preserve the conformal invariance thanks to the Einstein equation and the null energy condition.

So far, we do not know how to relax the gauge condition.\footnote{In the earlier version of the paper, we made a sign error which seems to invalidate the proof  presented there without using the gauge condition.} It would be interesting to see how the use of flux equations of motion may give additional constraint, and hopefully prove the general validity of the gauge condition.

To see the effect of the most general $h_i(\xi)$, we first note that by using the diffeomorphism invariance, we can always set $f(\xi) = g(\xi)$, and $h_i(\xi) = f(\xi) \tilde{h}_i(\xi)$. Since the symmetry is slightly reduced, it is no longer true that $R_{zz} + R_{tt}$ identically vanishes. However, with a suitable positive definite integrable factor, we can make it a total derivative:
\begin{align}
z^2\left( R_{zz} + R_{tt} \right) =  F(\xi) D^i j_i + G(\xi) (\partial_i  \tilde{h}_j - \partial_j \tilde{h}_i)^2 +H(\xi)\tilde{h}_i\tilde{h}^i \ ,  \label{totalR} \ 
\end{align}
where $D_i$ is the covariant derivative constructed from the internal metric $ds^{2}_{D-3}$. $F(\xi)$, $G(\xi)$ and $H(\xi)$ are certain positive definite functions of $\xi$, and $j_i$ is a certain local current made out of $h_i(\xi)$ and $f(\xi)$.\footnote{Explicitly they are given by: $F(\xi) = \frac{c_F}{f^{1/2} (1-f\tilde{h}_i\tilde{h}^i)^{3/2}}$, $G(\xi) = \frac{c_Gf^2}{(1-f\tilde{h}_i\tilde{h}^i)^2}$, $H(\xi) = \frac{c_H f}{(1-f\tilde{h}_i\tilde{h}^i)}$, and $j_i = \frac{c_j f^{3/2} \tilde{h}_i}{(1-f\tilde{h}_i\tilde{h}^i)^{1/2}}$. The constant factors depend on the dimensionality $D$.} 

We can integrate the equation over the internal space, and the left hand side is semi-positive definite due to the null energy condition. Then, the non-trivial $\tilde{h}_i$ can be seen as the effective violation of the null energy condition. It would be interesting to see whether the other equations motion make $\tilde{h}_i$ must vanish up to a diffeomorphism invariance. At this point, we temporarily conclude that the violation of Polchinski's theorem is only possible with the non-trivial KK vector fields $\tilde{h}_i$ is excited.

\section{In higher dimensions}
Originally,  Polchinski's theorem was proved only in $(1+1)$ dimension. However, since there is no known counterexamples in higher dimensions, it may be natural to generalize the conjecture in higher dimensions. Can we state a similar conjecture of the gravitational counterpart of Polchinski's theorem in higher dimensions and verify the conjecture in the supergravity limit? 

The conjecture itself can be easily generalized in higher dimensions. For definiteness, let us state the conjecture in the M-theory compactification. The string theory formulation will be straightforward.

{\bf Higher dimensional conjecture (in M-theory)}

The most general M-theory background ansatz that is  dual to a scale invariant $(1+d)$ dimensional field theory is given by
\begin{align}
ds^2_{11} = f(\xi) \frac{dz^2}{z^2} + g(\xi)\frac{-dt^2+dx_a^2}{z^2} + 2h_i(\xi) \frac{dz d\xi^i}{z} + ds^{2}_{9-d}(\xi) \ ,
\end{align}
where $a = 1,\cdots d$ and $i = d+1 \cdots 10$.
Similarly, the most general flux ansatz is given by
\begin{align}
F_4 = A + \frac{dz}{z} \wedge B + \frac{dt\wedge dx_1\wedge \cdots \wedge dx_d}{z^{d+1}} \wedge C + \frac{dz\wedge dt \wedge dx_1 \wedge \cdots \wedge dx_d}{z^{d+2}} \wedge D  \ ,
\end{align}
where $A$, $B$, $C$ and $D$ are differential forms on the internal manifold $X_{9-d}$ and do not depend on $(z,t,x_i)$. Of course, when $d$ is large enough, $C=D=0$ automatically.

The conjecture is that all the consistent solutions of M-theory that are dual  to a scale invariant field theory satisfy the conformal invariant condition: $f(\xi) = g(\xi)$, and $h_{i}(\xi) = B=C= 0$ up to diffeomorphism invariance.

{\bf Higher dimensional evidence in supergravity limit}

While there is no proof of the higher dimensional versions of Polchinski's theorem from the field theory, we can verify the conjecture in the supergravity limit  as in the previous section. Under the same gauge condition $h_i = 0$, we can still show
\begin{align}
 \frac{\alpha z^2}{F(\xi)} \left( g(\xi) R_{zz} + f(\xi) R_{tt} \right) = D^i j_i
\end{align}
where $D_i$ is the covariant derivative constructed from the internal metric $ds^{2}_{9-d}$, $F(\xi)$ is a positive function on $X_{D-3}$ and $j_i$ is a certain local current made out of $f(\xi)$ and $g(\xi)$.
Then by repeating the same argument in the previous section, and, in particular, by using the null energy condition of the supergravity action, one can verify that the higher dimensional conjecture holds in the supergravity limit (with no source): $h_{i}(\xi) = B=C= 0$ under the gauge condition.\footnote{The gauge condition reminds us of the compensator in strongly warped flux compactification: only when the compensator can be gauged away, the explicit computation of the Kahler potential is feasible.}
The verification of the conjecture shown here from the gravity computation may yield an evidential support for Polchinski's theorem in higher dimensions.

Within our arguments above, it is not so clear how the dimensionality of the non-compact space would affect the verification of the conjecture.
From the field theory perspective, $(1+1)$ dimension seems special because the tensor structure of the energy-momentum tensor is so restrictive, but there seems no apparent difference among higher dimensional compactifications in our discussion. On the contrary, the dimension of the compact space matters because if the compact dimension is one-dimension, for instance, the gauge condition is automatically true.

A similar situation has occurred in the gravitational counterpart of Zamolodchikov's c-theorem \cite{Zamolodchikov:1986gt} studied in \cite{Freedman:1999gp}. While the field theory argument is only valid in $(1+1)$ dimension, the gravity counterpart can be proved in any dimensions (as long as the energy momentum tensor satisfies a null energy condition). The dimensionality of the target space has not played any role there. However, given that the energy condition is {\it not} proved in the string/M-theory compactification, we can still make a non-trivial conjecture: the null energy condition is always true in the compactification with $(1+2)$ dimensional non-compact space because otherwise c-theorem is violated.\footnote{Indeed, the discussion here has led to a conjecture on the (non-)existence of a certain class of Sasaki-Einstein manifolds \cite{Gauntlett:2006vf}\cite{Nakayama:2007sb}, which can be proved mathematically.}
 With this regard, it would be interesting to observe that the positivity of the flux contribution of the energy momentum tensor, which is crucial in our derivation of gravitational Polchinski's theorem, is nothing but the null energy condition, so the assumption coincides with the gravitational c-theorem.
Note also that the proof of c-theorem and Polchinski's theorem parallels in $(1+1)$ dimension in the boundary field theory side \cite{Polchinski:1987dy}. 

We know that the violation of strong energy condition (which indicates the null energy condition) is not totally impossible in the string/M-theory: the possibility of inflation and the de-sitter space suggests that the strong energy condition must be violated. It is, however, important to note that even if the strong energy  condition (or more precisely null energy condition) were violated (with quantum corrections, brane sources, orientifold etc), the scale invariant but non-conformal field configuration would be strictly forbidden in $(1+2)$ dimension.
As we have seen, such holographic constraints on the flux compactification of the string/M-theory are quite powerful. In the next section, we will discuss the constraint on higher derivative/loop corrections in the string/M-theory from the holography.

\section{Constraint on higher derivative terms}
As we mentioned earlier, our conjecture is not restricted in the supergravity limit. Rather it should be applied to the full solution of the string/M-theory with all the higher derivative/loop corrections involved. The claim is that such a higher derivative/loop corrections conspire themselves so that the scale invariant but non-conformal field configurations are forbidden in the full string/M-theory. This will give us a highly non-trivial constraint on the higher-derivative/loop corrections. 

The ``spontaneous Lorentz symmetry breaking model" and the ``(gauged) ghost condensation model" studied in \cite{Nakayama:2009qu} are typical examples of effective action excluded from this argument. They realize the scale invariant but non-conformal field configurations so that such theories are inconsistent from the holographic viewpoint (unless we exclude such solutions from boundary/initial conditions).

The constraint on the higher derivative/loop corrections to the effective action from our conjecture is an interesting consistency check of the effective field theories.
As a typical example of higher derivative terms in the string theories, let us consider the DBI action coupled with the Einstein gravity:
\begin{align}
 S = \frac{1}{2\kappa^2} \int d^D x \sqrt{-g} R + \int d^{D}x \sqrt{-\mathrm{det} (g_{\mu\nu} + F_{\mu\nu}) } \ .
\end{align}
As in section 3, the coupling to the dilaton as well as the Chern-Simons terms do not change the argument, so we simply omit them in the following discussions.

We consider the background that is dual to a scale invariant field theory: the most general ansatz for the metric is given by
\begin{align}
ds^2_{D} = f(\xi) \frac{dz^2}{z^2} + g(\xi)\frac{-dt^2+dx^2}{z^2} + 2h_i(\xi) \frac{dz d\xi^i}{z} + ds^{2}_{D-3}(\xi) \ ,
\end{align}
while the most general ansatz for the DBI flux is given by
\begin{align}
F_2 = A + \frac{dz}{z} \wedge B + \frac{dt \wedge dx}{z^2} C.
\end{align}
Hereafter we impose the gauge condition $h_i(\xi) = 0$ to make the discussion simpler while keeping $f(\xi) \neq g(\xi)$ to maintain a certain generality.

The combination of the Ricci tensor: $g(\xi)R_{zz} + f(\xi) R_{tt}$ is proportional to a total derivative:
\begin{align}
 \frac{\alpha z^2}{f^{\frac{1}{2}}} \left( g(\xi) R_{zz} + f(\xi) R_{tt} \right) = D^i \left( -2f^{-\frac{1}{2}} g \partial_i f + 2 f^{\frac{1}{2}} \partial_i g \right) \ , \label{total}
\end{align}
where $D_i$ is the covariant derivative constructed from the internal metric $ds^{2}_{D-3}$, and $\alpha$ is a positive constant, which depends on the dimensionality of the space-time.
We see that the Einstein equation gives us semi-positive definite contributions to the left hand side as can be seen from the form of the DBI energy-momentum tensor:
\begin{align}
T_{\mu\nu} = \frac{\sqrt{\mathrm{det}(g_{\mu\nu} + F_{\mu\nu})}}{2\sqrt{\mathrm{det}g_{\mu\nu}}} \left[(g_{\mu\nu} + F_{\mu\nu})^{-1} +(g_{\nu\mu} + F_{\nu\mu})^{-1}\right] \ ,
\end{align}
so 
\begin{align}
\kappa^2 z^2\left( g(\xi)R_{zz} + f(\xi) R_{tt}\right) = F(\xi) \left(\frac{1}{1-\frac{C^2}{g^2}} - \frac{1}{1+ \frac{B_i(g_{ij}+A_{ij})^{-1}B_j}{f^2}}\right) \ .
\end{align}
with a positive definite function $F(\xi)$.
By integrating it over the internal space, we see that $B = C= 0$, and hence $f(\xi) = g(\xi)$ from the $R_{zi}$ component of Einstein's equation. In this way, the class of solutions of the DBI action coupled with gravity studied here are guaranteed to be consistent with Polchinski's theorem. However, if the kinetic term of the gauge field were generically given $ S = \int d^Dx \sqrt{-g} \mathcal{L}(F_{\mu\nu}, g_{\mu\nu})$,  the positivity of $g(\xi) R_{zz} + f(\xi)R_{tt}$ from the Einstein equation would  no longer follow automatically, and there could be scale invariant but non-conformal field configurations as solutions of the equations of motion. If we could find such a solution within a certain higher derivative action, then we would have to discard the theory as it predicts an inconsistent field configuration with holography like the models studied in \cite{Nakayama:2009qu}.

In a similar manner, the higher derivative corrections of the string/M-theory must be compatible with the conjecture presented in this paper. It would be interesting to investigate its structure and the power of the holographic constraint further. The discussion in this section reminds us of the constraint of the higher derivative terms coming from the superluminal propagation studied in \cite{Adams:2006sv}. Since both of the discussions somehow involve the spontaneous Lorentz symmetry breaking, there might be a deep connection. Our conjecture is totally based on the holographic argument, however, so it is very interesting to understand how the effective field theories would  become inconsistent if one studies the scale invariant but non-conformal configurations directly from the gravity viewpoint.

\section{Discussion}
In this paper, we have discussed how the gravitational counterpart of Polchinski's theorem leads to a no-go condition on some field configurations. In order for the conjecture to be true in the string/M-theory, the structure of the higher derivative terms must be constrained. Our criterion can be used as a tool to judge whether a given theory is consistent as a quantum theory of gravity.
Here, we would like to discuss possible loopholes in the argument before we conclude the paper.

First of all, there is a logical possibility that the holography is wrong. We cannot exclude this possibility while we believe that it is quite unlikely given the success of holographic computations of entropy, AdS/CFT correspondence etc. In particular, when the cosmological constant is positive, there is no rigorous definition of holography (while there are many proposals in literatures e.g. \cite{Strominger:2001pn}\cite{Freivogel:2006xu}\cite{Sekino:2009kv}). The background studied in this paper has a negative cosmological constant, so it is unlikely that the boundary theory is not well-defined. Indeed, we could always operationally define the boundary theory by using the holographic prescription of the correlation functions \cite{Gubser:1998bc}\cite{Witten:1998qj}.\footnote{However, we have to still check whether the correlation functions satisfy the axiom of boundary field theories. This has been recently studied in \cite{Heemskerk:2009pn}.} In addition, our study shows that in the supergravity limit (under a certain gauge condition), the string/M-theory is consistent with the holographic Polchinski's theorem.

The second possibility is that one of the assumptions in Polchinski's theorem is violated. In particular, the gravity theory that is dual to a scale invariant but non-conformal field theory might be non-unitary. This is indeed possible: in our example, if we could flip the sign of the kinetic terms of flux by hand, we would find a non-unitary scale invariant but non-conformal configurations. Of course, this should not happen in any consistent theories of quantum gravity: unitarity is the most fundamental principle of quantum mechanics, and we do not want to abandon it in any way. Thus, even if there existed scale invariant but non-conformal field configurations due to the loss of unitarity, such a theory would be equally inconsistent and should be forbidden. Again, we have shown  the string/M-theory is consistent with the holographic Polchinski's theorem with the unitarity assumption.

The third possibility is that even if the effective action would admit such scale invariant but non-conformal field configurations, we might forbid such configurations by imposing boundary (initial) conditions. Whether such a condition is a dynamical one or (artificially) given one is a delicate question. For instance, Einstein's gravity as it is admits a solution with a time-like closed curve, but whether such a solution should be forbidden dynamically or by an initial condition is a subtle issue. Given the success of the holographic c-theorem and the supergravity limit of our conjecture, however, we are inclined to believe that the scale invariant but non-conformal field configurations are forbidden at the action level, rather than by the initial (boundary) condition. 

We cannot exclude all these possibilities, yet it is clear that any of the possibilities, as well as our original proposal in the main part of the paper, would be exciting of its own to uncover the landscape and swampland in the string/M-theory. We hope that the further holographic study of the map will facilitate our navigation in search of the consistent quantum theories of gravity to reach the ultimate theory of the universe.

\section*{Acknowledgements}
The work was supported in part by the National Science Foundation under Grant No.\ PHY05-55662 and the UC Berkeley Center for Theoretical Physics. The author would like to thank YITP for their hospitality, where a part of this work was done.

\end{document}